\documentclass[prd,onecolumn,nofootinbib,showkeys]{revtex4}
\usepackage{amssymb,amsmath,bm}

\begin{document}

\title{BIG BOUNCE FROM SPIN AND TORSION}

\author{{\bf Nikodem J. Pop{\l}awski}}

\affiliation{Department of Physics, Indiana University, Swain Hall West, 727 East Third Street, Bloomington, Indiana 47405, USA}
\email{nipoplaw@indiana.edu}

\noindent
{\em General Relativity and Gravitation}\\
Vol. {\bf 44}, No. 4 (2012) pp. 1007--1014\\
\copyright\,Springer Science+Business Media, LLC
\vspace{0.4in}

\begin{abstract}
The Einstein-Cartan-Sciama-Kibble theory of gravity naturally extends general relativity to account for the intrinsic spin of matter.
Spacetime torsion, generated by spin of Dirac fields, induces gravitational repulsion in fermionic matter at extremely high densities and prevents the formation of singularities.
Accordingly, the big bang is replaced by a bounce that occurred when the energy density $\epsilon\propto gT^4$ was on the order of $n^2/m_\textrm{Pl}^2$ (in natural units), where $n\propto gT^3$ is the fermion number density and $g$ is the number of thermal degrees of freedom.
If the early Universe contained only the known standard-model particles ($g\approx 100$), then the energy density at the big bounce was about 15 times larger than the Planck energy.
The minimum scale factor of the Universe (at the bounce) was about $10^{32}$ times smaller than its present value, giving $\approx 50\,\mu\mbox{m}$.
If more fermions existed in the early Universe, then the spin-torsion coupling causes a bounce at a lower energy and larger scale factor.
Recent observations of high-energy photons from gamma-ray bursts indicate that spacetime may behave classically even at scales below the Planck length, supporting the classical spin-torsion mechanism of the big bounce.
Such a classical bounce prevents the matter in the contracting Universe from reaching the conditions at which a quantum bounce could possibly occur.
\end{abstract}


\maketitle

The Einstein-Cartan-Sciama-Kibble (ECSK) theory of gravity naturally extends Einstein's general relativity (GR) to account for the quantum-mechanical, intrinsic angular momentum (spin) of elementary particles that compose gravitating matter \cite{KS1,KS2,KS3,KS4,Hehl1,Hehl2,Hehl3,Hehl4,Hehl5,Hehl6,rev,Lo}.
The ECSK gravity is based on the Lagrangian density for the gravitational field that is proportional to the curvature scalar, as in GR \cite{LL}.
It removes, however, the constraint of GR that the torsion tensor (the antisymmetric part of the affine connection) be zero by promoting this tensor to a dynamical variable, as the metric tensor \cite{KS1,KS2,KS3,KS4,Hehl1,Hehl2,Hehl3,Hehl4,Hehl5,Hehl6,rev,Lo}.
The torsion tensor is then given by the principle of least action and in many physical situations it turns out to be zero.
But in the presence of fermions, which compose all stars in the Universe, spacetime torsion does not vanish because Dirac fields couple minimally to the torsion tensor \cite{KS1,KS2,KS3,KS4,Hehl1,Hehl2,Hehl3,Hehl4,Hehl5,Hehl6,rev,Lo}.
At macroscopic scales, such particles can be averaged and described as a spin fluid \cite{spin_fluid1,spin_fluid2,avert_avg}.
It has been shown in \cite{NSH} that the spin-fluid form of the spin tensor results from the conservation law for this tensor \cite{Hehl1,Hehl2,Hehl3,Hehl4,Hehl5,Hehl6,rev,Lo}.

The field equations of the ECSK gravity can be written as the general-relativistic Einstein equations with the modified energy-momentum tensor \cite{KS1,KS2,KS3,KS4,Hehl1,Hehl2,Hehl3,Hehl4,Hehl5,Hehl6,rev,Lo}.
Such a tensor has terms which are quadratic in the spin tensor and thus do not vanish after averaging \cite{avert_avg,Kuch}.
These terms are significant only at densities of matter that are much larger than the density of nuclear matter; otherwise the ECSK gravity effectively reduces to GR.
The ECSK gravity therefore passes all current tests of GR.
These terms generate gravitational repulsion in spin-fluid fermionic matter, which becomes significant in the early Universe and inside black holes.
Such a repulsion prevents the formation of singularities from fermionic matter \cite{avert_avg,Kuch,avert1,avert2,avert3,avert4}.
It replaces the singular big bang by a nonsingular state of minimum but finite radius \cite{Kuch,Gas,infl1,infl2}.
This extremely hot and dense state is a (big) bounce that follows a contracting phase of the Universe and initiates its rapid expansion \cite{bounce1,bounce2,bounce3,bounce4,bounce5,bounce6,bounce7,bounce8,bounce9}.

In \cite{infl1,infl2}, we considered the dynamics of a closed universe immediately after such a bounce.
We showed that a negative and extremely small (in magnitude) spin-torsion density parameter naturally explains why the present Universe appears spatially flat, homogeneous, and isotropic.
The ECSK gravity therefore not only eliminates an unphysical cosmological singularity but also provides an alternative to cosmic inflation without requiring additional fields and specific assumptions on their potentials.
Another advantage of the ECSK theory is that it has no free parameters.
We also suggested that the coupling between spin and torsion may be the mechanism that allows for a scenario in which every black hole produces a new universe inside, instead of a singularity \cite{infl1,infl2}.
The contraction of our Universe before the bounce at the minimum radius may thus correspond to the dynamics of matter inside a collapsing black hole existing in another universe \cite{BH1,BH2,BH3,BH4,BH5,BH6,BH7,BH8,BH9}.
A scenario in which the Universe was born in a black hole seems more reasonable than its contraction from infinity in the past \cite{past} because the latter does not explain what caused such a contraction.
If our Universe was born in a black hole that has formed in a parent universe, then it would interact with the parent universe.
Recent measurements of the large-scale bulk flows of galaxy clusters \cite{Kash}, which cannot be explained within the standard theoretical framework, suggest that our Universe may be interacting with other parts of spacetime.
Torsion may therefore provide a natural scenario for what existed before the Universe began to expand \cite{infl1,infl2,BH1,BH2,BH3,BH4,BH5,BH6,BH7,BH8,BH9}.

In \cite{infl1,infl2}, we estimated the torsion density parameter and the conditions at the big bounce from the current number density of neutrinos which are the most abundant fermions in the Universe. 
We found that the energy density of matter at the bounce is a few orders of magnitude larger than the Planck energy density.
The scenario of a big bounce within the classical ECSK theory may therefore be inadequate because the Planck regime is expected to be described by a quantum theory of gravity.
Interestingly, loop quantum gravity (LQG), which assumes that spacetime is discrete at the Planck scale, also predicts a cosmic bounce at the Planck energy \cite{LQG1,LQG2,LQG3}.
In this paper, we refine the results of \cite{infl1,infl2} by including the thermal degrees of freedom arising not only from photons and neutrinos, but also from other standard-model particles that are ultrarelativistic in the early Universe.

We consider a closed ($k=1$), homogeneous, and isotropic universe filled with fermionic matter macroscopically averaged as a spin fluid \cite{avert_avg}.
The Einstein-Cartan field equations for the Friedman-Lema\^{i}tre-Robertson-Walker (FLRW) metric describing such a universe are given by the Friedman equations for the scale factor $a(t)$ \cite{infl1,infl2,Kuch,Gas}:
\begin{eqnarray}
& & \frac{1}{c^2}\Bigl(\frac{da}{dt}\Bigr)^2+k=\frac{1}{3}\kappa\Bigl(\epsilon+\epsilon_\textrm{S}\Bigr)a^2+\frac{1}{3}\Lambda a^2, \label{Fri1} \\
& & \frac{d}{dt}\bigl((\epsilon+\epsilon_\textrm{S})a^3\bigr)+(p+p_\textrm{S})\frac{d}{dt}(a^3)=0,
\end{eqnarray}
where
\begin{equation}
\epsilon_\textrm{S}=-\frac{1}{4}\kappa s^2,\,\,\,p_\textrm{S}=\epsilon_\textrm{S}
\label{effective}
\end{equation}
are the contributions to the energy density of matter $\epsilon$ and pressure $p$ from the spin-torsion coupling \cite{avert_avg}.
The quantity $s^2$ in (\ref{effective}) is the square of the dispersion of the spin density distribution around its average value and it is equal, for unpolarized spins, to \cite{NP}
\begin{equation}
s^2=\frac{1}{8}(\hbar cn)^2,
\end{equation}
where $n$ is the fermion number density.
In the early Universe, the matter is ultrarelativistic: $p\approx\epsilon/3$.
We can also neglect in (\ref{Fri1}) the terms with $k$ and the cosmological constant $\Lambda$.
The big bounce occurs when $da/dt=0$, which gives
\begin{equation}
\epsilon=\frac{1}{4}\kappa s^2=\frac{1}{32}\kappa(\hbar cn)^2.
\label{zero}
\end{equation}
In natural units ($\hbar=c=1$), this condition is
\begin{equation}
\epsilon=\frac{\pi}{4}\frac{n^2}{m_\textrm{Pl}^2},
\end{equation}
where $m_\textrm{Pl}$ is the Planck mass.

The energy density of ultrarelativistic matter in kinetic equilibrium is given by \cite{cosmo1,cosmo2}
\begin{equation}
\epsilon(T)=\frac{\pi^2}{30}g_\star(T)\frac{(k_\textrm{B}T)^4}{(\hbar c)^3},
\label{energy}
\end{equation}
where $T$ is the temperature of the early Universe.
The effective number of thermal degrees of freedom $g_\star(T)$ is given by \cite{cosmo1,cosmo2}
\begin{equation}
g_\star(T)=g_\textrm{b}(T)+\frac{7}{8}g_\textrm{f}(T),
\end{equation}
where $g_\textrm{b}=\sum_i g_i$ is summed over relativistic bosons and $g_\textrm{f}=\sum_i g_i$ is summed over relativistic fermions.
For each particle species, $g_i$ is the number of its spin states.
We include in $g_\star(T)$ only relativistic species (whose rest masses $m_i$ satisfy 
$m_i c^2<k_\textrm{B}T$) because the energy density of relativistic particles in the early Universe is much larger than that of nonrelativistic particles.
The fermion number density is given by \cite{cosmo1,cosmo2}
\begin{equation}
n(T)=\frac{\zeta(3)}{\pi^2}g_n(T)\frac{(k_\textrm{B}T)^3}{(\hbar c)^3},
\label{number}
\end{equation}
where $\zeta(3)\approx 1.202$ is the Riemann zeta function of 3, and
\begin{equation}
g_n(T)=\frac{3}{4}g_\textrm{f}(T).
\end{equation}

Substituting the energy density (\ref{energy}) and the number density (\ref{number}) into the condition for a torsion-driven bounce (\ref{zero}) gives the temperature at the big bounce:
\begin{equation}
T_\textrm{bb}=\biggl(\frac{64}{270}\biggr)^{1/2}\frac{\pi^{5/2}}{\zeta(3)}\frac{(g_\textrm{b}+7g_\textrm{f}/8)^{1/2}}{g_\textrm{f}}T_\textrm{Pl},
\label{tb}
\end{equation}
where $T_\textrm{Pl}=k_\textrm{B}^{-1}(\hbar c^5/G)^{1/2}$ is the Planck temperature.
Substituting the temperature (\ref{tb}) into (\ref{energy}) gives the energy density of matter at the bounce:
\begin{equation}
\epsilon_\textrm{bb}=\xi\frac{(g_\textrm{b}+7g_\textrm{f}/8)^3}{g_\textrm{f}^4}\epsilon_\textrm{Pl},
\label{eb}
\end{equation}
where $\epsilon_\textrm{Pl}=c^7/(\hbar G^2)=5.1\times10^{96}\mbox{kg}/\mbox{m}^3\cdot c^2$ is the Planck energy density and
\begin{equation}
\xi=\frac{1}{30}\biggl(\frac{64\pi^6}{270\zeta^2(3)}\biggr)^2.
\end{equation}
For $T>m_t$, where $m_t\approx\,175\,\mbox{GeV}$ is the rest mass of a $t$ quark, all known particles are relativistic.
If the early Universe contained only all known standard-model particles (with equal temperatures), then $g_\textrm{b}=28$ and $g_\textrm{f}=90$ \cite{cosmo1,cosmo2} at the big bounce, which gives
\begin{equation}
\epsilon_\textrm{bb}=1.54\times10^1\,\epsilon_\textrm{Pl}.
\end{equation}
In this case, the energy density at the bounce is at the Planck scale, where the classical ECSK theory should be replaced by a quantum theory of gravity.
LQG predicts a quantum cosmic bounce at the same scale \cite{LQG1,LQG2,LQG3}.
If, however, much more fermionic degrees of freedom existed at extremely high energies, then the spin-torsion coupling causes a bounce below the Planck energy.
Such a scenario would be possible, for example, if standard-model fermions were composed of more elementary particles \cite{preon1,preon2,preon3,preon4,preon5,preon6}.
If, for example, $g_f=10^5$ at the bounce and the ratio $g_\textrm{b}/g_\textrm{f}$ is constant, then $\epsilon_\textrm{bb}=1.4\times10^{-2}\,\epsilon_\textrm{Pl}$.
In this case, the classical description of gravity is sufficient.
Such a classical bounce would prevent the matter in the contracting Universe from reaching densities at which a quantum bounce would happen (if LQG is correct).
Consequently, LQG would not be able to provide cosmological signatures in the presence of torsion.

Recent observations of high-energy photons from gamma-ray bursts \cite{obs1,obs2} suggest, however, that spacetime may be continuous at the Planck length and lower scales \cite{contin}.
Spacetime may therefore behave classically even above the Planck energy.
As a result, the classical spin-torsion mechanism of the big bounce should be valid without additional fermionic degrees of freedom.

The energy density of ultrarelativistic matter scales according to $\epsilon_\textrm{R}\sim a^{-4}$, so that its present-day value is given by
\begin{equation}
\epsilon_{\textrm{R}0}=\epsilon_\textrm{bb}\biggl(\frac{a_\textrm{bb}}{a_0}\biggr)^4\frac{(g_\textrm{b}+7g_\textrm{f}/8)|_0}{(g_\textrm{b}+7g_\textrm{f}/8)|_\textrm{bb}},
\label{scale1}
\end{equation}
where $a_\textrm{bb}$ is the scale factor at the bounce and $a_0$ is the current scale factor.
Subscripts $0$ denote quantities evaluated at the present time.
The second ratio on the right-hand side of (\ref{scale1}) represents the decrease of the number of relativistic degrees of freedom: their present-day values are $g_\textrm{b}=2$ (photon) and $g_\textrm{f}=6$ (neutrinos and antineutrinos) \cite{cosmo1,cosmo2}.
Using the current density parameter for radiation, $\Omega_\textrm{R}=\epsilon_{\textrm{R}0}/\epsilon_\textrm{cr}$, where $\epsilon_\textrm{cr}=9.24\times10^{-27}\mbox{kg}/\mbox{m}^3\cdot c^2$ is the current critical energy density \cite{infl1,infl2,WMAP}, we obtain
\begin{equation}
\frac{a_\textrm{bb}}{a_0}=\biggl(\frac{\Omega_\textrm{R}\epsilon_\textrm{cr}}{\epsilon_\textrm{bb}}\frac{(g_\textrm{b}+7g_\textrm{f}/8)|_\textrm{bb}}{(g_\textrm{b}+7g_\textrm{f}/8)|_0}\biggr)^{1/4}.
\label{value01}
\end{equation}
Putting $\Omega_\textrm{R}=8.8\times10^{-5}$ \cite{infl1,infl2,WMAP} gives
\begin{equation}
\frac{a_\textrm{bb}}{a_0}=2\times10^{-32}.
\label{value1}
\end{equation}

The spin-torsion contribution to the energy density, $\epsilon_\textrm{S}\propto n^2$, scales according to $\epsilon_\textrm{S}\sim a^{-6}$ \cite{Kuch,Gas,infl1,infl2}, so that its present-day value is given by
\begin{equation}
\epsilon_{\textrm{S}0}=\epsilon_\textrm{S}|_\textrm{bb}\biggl(\frac{a_\textrm{bb}}{a_0}\biggr)^6\biggl(\frac{g_\textrm{f}|_0}{g_\textrm{f}|_\textrm{bb}}\biggr)^2.
\label{scale2}
\end{equation}
The second ratio on the right-hand side of (\ref{scale2}) represents the decrease of the number of relativistic degrees of freedom.
The condition (\ref{zero}) reads
\begin{equation}
\epsilon_\textrm{S}|_\textrm{bb}=-\epsilon_\textrm{bb}.
\end{equation}
Using the current density parameter for the spin-torsion coupling, $\Omega_\textrm{S}=\epsilon_{\textrm{S}0}/\epsilon_\textrm{cr}$ \cite{infl1,infl2}, we obtain
\begin{equation}
\frac{a_\textrm{bb}}{a_0}=\biggl(-\frac{\Omega_\textrm{S}\epsilon_\textrm{cr}}{\epsilon_\textrm{bb}}\frac{g^2_\textrm{f}|_\textrm{bb}}{g^2_\textrm{f}|_0}\biggr)^{1/6}.
\label{value02}
\end{equation}
Putting $\Omega_\textrm{S}=-8.6\times10^{-70}$ \cite{infl1,infl2} gives
\begin{equation}
\frac{a_\textrm{bb}}{a_0}=1.7\times10^{-32}.
\label{value2}
\end{equation}
This value is smaller than (\ref{value1}) because $\Omega_\textrm{R}$, which appears in (\ref{value01}) in the $(1/4)$-th power, is increased due to pair annihilation (such as electron-positron annihilation) into photons \cite{cosmo1,cosmo2}.

Eliminating $\epsilon_\textrm{cr}/\epsilon_\textrm{bb}$ from (\ref{value01}) and (\ref{value02}) yields
\begin{equation}
\frac{a_\textrm{bb}}{a_0}=\sqrt{-\frac{\Omega_\textrm{S}}{\Omega_\textrm{R}}}\frac{g_\textrm{f}|_\textrm{bb}}{g_\textrm{f}|_0}\biggl(\frac{(g_\textrm{b}+7g_\textrm{f}/8)|_0}{(g_\textrm{b}+7g_\textrm{f}/8)|_\textrm{bb}}\biggr)^{1/2}.
\label{value03}
\end{equation}
This relation refines the formula for the minimum normalized scale factor $a_\textrm{bb}/a_0=\sqrt{-\Omega_\textrm{S}/\Omega_\textrm{R}}$ found in \cite{infl1,infl2}.
Putting the values of $\Omega_\textrm{R}$ and $\Omega_\textrm{S}$ \cite{infl1,infl2} gives
\begin{equation}
\frac{a_\textrm{bb}}{a_0}=1.2\times10^{-32}.
\label{value3}
\end{equation}
This value is smaller than (\ref{value2}) because $\Omega_\textrm{R}$ appears in (\ref{value03}) in the $(-1/2)$-th power.
Since neutrinos barely interact with matter, the value (\ref{value02}) is not affected by pair annihilation and it should be more accurate than (\ref{value01}) and (\ref{value03}).
Accordingly, the normalized scale factor at the big bounce is equal to (\ref{value2}).
Using the present-day scale factor $a_0=2.9\times10^{27}\mbox{m}$ \cite{infl1,infl2}, derived from the WMAP data \cite{WMAP}, therefore gives the value of the scale factor of the Universe at its minimum size (at the bounce):
\begin{equation}
a_\textrm{bb}=4.9\times10^{-5}\mbox{m}.
\end{equation}

Eliminating $a_\textrm{bb}/a_0$ from (\ref{value01}) and (\ref{value02}), and using (\ref{eb}) leads to
\begin{equation}
\Omega_\textrm{S}=-\sqrt{\frac{\Omega^3_\textrm{R}}{\Omega_\textrm{Pl}}}\biggl(\frac{g_\textrm{f}^4|_0}{\xi(g_\textrm{b}+7g_\textrm{f}/8)^3|_0}\biggr)^{1/2},
\end{equation}
where $\Omega_\textrm{Pl}=\epsilon_\textrm{Pl}/\epsilon_\textrm{cr}$.
This relation shows why the spin-torsion density parameter $\Omega_\textrm{S}$ is extremely small in magnitude, which can explain the flatness and horizon problems without inflation \cite{infl1,infl2}.
$|\Omega_\textrm{S}|\ll1$ is caused by an extremely large $\Omega_\textrm{Pl}=5.5\times10^{122}$.

\section*{Acknowledgments}
The author would like to thank Brian Fields for helpful remarks regarding the thermal history of the Universe, Shantanu Desai for the hospitality at the University of Illinois at Urbana-Champaign, and the Office of Postdoctoral Affairs at Indiana University Bloomington for a travel award.

\end{document}